\documentclass{article}
\usepackage{amssymb}
\usepackage{amsmath}
\usepackage{amsfonts}
\usepackage{amsthm}
\usepackage[T1]{fontenc}
\usepackage{authblk}

\newtheorem{definition}{Definition}
\newtheorem{proposition}{Proposition}

\newtheorem{remark}{Remark}

\begin{document}
\title{Toric data and Killing forms on homogeneous Sasaki-Einstein manifold $T^{1,1}$}

\author[1]{Vladimir Slesar\thanks{vlslesar@central.ucv.ro}}
\author[2]{Mihai Visinescu\thanks{mvisin@theory.nipne.ro}}
\author[3,4]{Gabriel Eduard
V\^ilcu\thanks{gvilcu@upg-ploiesti.ro}\thanks{gvilcu@gta.math.unibuc.ro}}
\affil[1]{Department of Mathematics, University of Craiova,

Str. Al.I. Cuza, Nr. 13, Craiova 200585, Romania}
\affil[2]{Department of Theoretical Physics,

National Institute for Physics and Nuclear Engineering,

Magurele, P.O.Box M.G.-6, Romania}
\affil[3]{Department of Cybernetics and Economic Informatics,

Petroleum-Gas University of Ploie\c sti,

Bd. Bucure\c sti, Nr. 39, Ploie\c sti 100680, Romania}
\affil[4]{Faculty of Mathematics and Computer Science,

Research Center in Geometry, Topology and Algebra,

University of Bucharest, Str. Academiei, Nr. 14, Sector 1,
Bucharest 060042, Romania}
\date{\today }
\maketitle

\begin{abstract}

Throughout this paper we investigate the complex structure of the
conifold $C(T^{1,1})$ basically making use of the interplay between symplectic
and complex approaches of the K\"{a}hler toric manifolds.
The description of the Calabi-Yau manifold $C(T^{1,1})$ using toric data
allows us to write explicitly the complex coordinates and apply standard
methods for extracting special Killing forms on the base manifold. As an outcome,
we obtain the  complete set of special Killing forms on the five-dimensional
Sasaki-Einstein space $T^{1,1}$.

\end{abstract}

\section{Introduction}

Symmetries are widely used as a useful tool in modeling physical
systems. The ordinary symmetries are associated with isometries, that
are spacetime diffeomorphisms that leave the metric invariant. A
one-parameter continuous isometry is connected with  a Killing
vector field. An extension of the Killing vector fields is
represented by conformal Killing vector fields \cite{K-Y} with flows
preserving a given class of metrics.

However, it has been proved that the investigation of symmetries in
the whole phase space of a system is exceedingly useful. Such
transformations of the whole phase space for which the dynamics
of the system is left invariant are often referred as {\it hidden
symmetries}. The hidden symmetries of curved manifolds are represented
by Killing tensors and Killing-Yano tensors. Thanks to such symmetries
many complicated physical problems become tractable taking into account
that the equation of motion are separable and integrable. Analogously,
conformal Killing-Yano tensors are associated with conserved quantities
along null geodesics  and integrability of massless field equations.

The purpose of this paper is to present a method to construct Killing
forms on toric Sasaki-Einstein manifolds. We exemplify the procedure in
the case of the five-dimensional homogeneous Sasaki-Einstein manifold
$T^{1,1}$.

Until recently the only explicitly known non-trivial
Sasaki-Einstein metric in dimension five  was $T^{1,1}$ \cite{C-O}.
The five-dimensional manifolds $T^{p,q}$ which are the coset spaces
$(SU(2)\times SU(2))/U(1)$ have been considered by Romans \cite{LJR}
in the context of Kaluza-Klein supergravity. Romans found that for
$p=q=1$ the compactification preserves $8$ supersymmetries, while for
other $p$ and $q$ all supersymmetries are broken.

In light of the $AdS/CFT$ correspondence the $AdS_5 \times T^{1,1}$
model of \cite{KW} is the first example of a supersymmetric holographic
theory based on a compact manifold which is not locally $S^5$.

The approach we take in order to achieve our goal basically relies on
the interplay between symplectic and complex coordinates on toric manifolds
\cite{M-S-Y,Abr}.
In our particular case, this description of the conifold is slightly
different from \cite{M-S,H-K-O}, when the geometric features of the
conifold are mainly exhibited. In turn, our approach enable us
to use the correspondence between special Killing forms and parallel
forms on the metric cone which was introduced by Semmelmann \cite{Semm}.
For more details concerning this method in the case of toric manifolds
we refer to \cite{Vis,S-V-V}.

The paper is organized in the following manner: In the second Section
we introduces the main concepts and technical tools we use in the
rest of the paper. We also give the necessary preliminaries regarding
the special Killing forms, toric Sasaki-Einstein manifolds and the
way their Calabi-Yau cones spaces can be regarded as complex manifolds.
In Section 3 we apply these results constructing complex coordinates
in the particular case of the conifold $C(T^{1,1})$. In Section 4 we
extract the  complete set of special Killing forms on the Sasaki-Einstein
space $T^{1,1}$. Finally, our conclusions are presented within the last
Section.

\section{Preliminaries}

\subsection{Special Killing forms}

A natural generalization of conformal Killing vector fields is given by
the conformal Killing forms which are sometimes referred as twistor
forms or conformal Killing-Yano tensors.
\begin{definition}
A conformal Killing-Yano tensor of rank $p$  on a $n$-dimensional
Riemannian manifold $(M,g)$ is a $p$-form $\psi$ which satisfies
\begin{equation}\label{CKY}
\nabla_X\psi=\frac{1}{p+1}X \lrcorner
d\psi-\frac{1}{n-p+1}X^*\wedge d^*\psi \,,
\end{equation}
for any vector field $X$ on $M$.
\end{definition}
Here we used the standard conventions: $\nabla$ is the Levi-Civita
connection with respect to the metric $g$, $X^*$ is the $1$-form dual
to the vector field $X$, $\lrcorner$ is the operator dual to the wedge
product and $d^*$ is the adjoint of the exterior derivative $d$.

In component notation, the conformal Killing-Yano tensor equation is
given by
\[
\nabla_{(i_1} \psi_{i_2)i_3 \cdots i_{p+1}} = \frac{1}{n-p+1}
\left( g_{i_1 i_2} \nabla_j \psi^{j}_{\phantom{j} i_3 \cdots i_{p+1}}
- (p-1)g_{[i_3(i_1} \nabla_j \psi^{j}_{\phantom{j} i_2)i_4
\cdots i_{p+1}]}\right)\,.
\]
We used round brackets to denote symmetrization over the indices
within. For $p=1$ we recover the usual definition of a Killing vector:
\[
\nabla_{(j}\psi_{i)} = 0 \,.
\]
If $\psi$ is co-closed in (\ref{CKY}), then we obtain the definition
of a Killing-Yano tensor \cite{K-Y} which, in component notation,
satisfies the equation:
\[
\nabla_{(j}\psi_{i_1)i_2 \dots i_p} = 0 \,.
\]

A particular class of Killing forms is represented by the special
Killing forms:
\begin{definition}
A Killing form $\psi$ is said to be a special Killing form if it
satisfies for some constant $c$ the additional equation
\[
\nabla_X(d\psi) = c X^* \wedge \psi \,,
\]
for any vector field $X$ on $M$.
\end{definition}

It is worth mentioning the fact that the most known Killing forms
are actually special.

There is also a symmetric generalization of the Killing vectors:
\begin{definition}
A symmetric tensor $K_{i_1 \cdots i_r}$ of rank $r>1$ satisfying
the generalized Killing equation
\[
\nabla_{(j}K_{i_1 \cdots i_r)} =0\,,
\]
is called a St\"{a}ckel-Killing tensor.
\end{definition}

The analogue of the conserved quantities associated with Killing
vectors is given by the following proposition:
\begin{proposition}
For any geodesic $\gamma$ with tangent vector $\dot{\gamma}^i$
\[
Q_K =K_{i_1 \cdots i_r} \dot{\gamma}^{i_1} \cdots \dot{\gamma}^{i_r}\,,
\]
is constant along $\gamma$.
\end{proposition}

Let us note that there is an important connection between these
two generalizations of the Killing vectors. To wit, given two
Killing-Yano tensors $\psi^{i_1, \dots, i_k}$ and
$\sigma^{i_1, \dots, i_k}$ there is a St\"{a}ckel-Killing tensor of
rank $2$:
\[
K^{(\psi,\sigma)}_{ij} = \psi_{i i_2 \dots i_k}
\sigma_{j}^{\phantom{j}i_2 \dots i_k}+ \sigma_{i i_2 \dots i_k}
\psi_{j}^{\phantom{j}i_2 \dots i_k} \,.
\]
This fact offers a method to generate higher order integrals of
motion by identifying the complete set  of Killing forms.

\subsection{Sasaki-Einstein manifolds}

\begin{definition}
An \emph{almost contact
structure}  on a smooth manifold  $M$ is a triple
$(\varphi,B,\eta)$, where $\varphi$ is a field  of endomorphisms
of the tangent spaces, $B$ is a vector field and $\eta$ is a
1-form on $M$ satisfying (see \cite{Sas})
\begin{equation}\label{gg1}
\varphi^2=-I+\eta\otimes B,\ \ \  \eta(B)=1 \,.
\end{equation}
\end{definition}

We remark that many authors also include in the above definition the
conditions that $\varphi B=0$ and $\eta\circ\varphi=0$, although these
are deducible from (\ref{gg1}) (see \cite{Bl}).

A Riemannian metric $g$ on $M$ is said to be \emph{compatible}
with the almost contact structure  $(\varphi,B,\eta)$ if and only if
the relation
\[
g(\varphi X, \varphi Y)=g(X,Y)-\eta(X)\eta(Y)\,,
\]
holds for all pairs of vector fields $X,Y$ on $M$. In this case
$(\varphi,B,\eta,g)$ is called an \emph{almost contact metric structure}.
Moreover, if the Levi-Civita
connection $\nabla$ of the metric $g$ satisfies
\[
(\nabla_X\varphi)
Y=g(X,Y)B-\eta(Y)X\,,
\]
for all vector fields $X,Y$ on $M$, then $(\varphi,B,\eta,g)$ is said to be a
\emph{Sasakian structure} \cite{Bl}.

It is also important to note that Sasakian geometry is in fact the
odd-dimensional counterpart of K\"{a}hler geometry, since
a Sasakian structure may be reinterpreted and characterized
in terms of the metric cone as follows.
The metric cone of a Riemannian manifold
$(M,g)$ is the Riemannian manifold $C(M)=(0,\infty)\times M$
with the metric given by
\[
\bar{g}=dr^2+r^2g\,,
\]
where $r$ is a
coordinate on $(0,\infty)$. Then $M$ is a Sasaki manifold if and only
if its metric cone $C(M)$ is  K\"{a}hler \cite{B-G-1999}. We note that the one form
$\eta$ extends to a one form on $C(M)$ by
$\eta(X)=\frac12\bar{g}(B,X)$ and $M$ is identified with the subset $r=1$ of $C(M)$. In
particular, the cone $C(M)$ is equipped with an integrable
complex structure $J$ defined by
\[
Jr\partial_r=B,\ \ JY=\varphi Y-\eta(Y)r\partial_r,\ Y\in TM \,,
\]
and a K\"{a}hler 2-form $\omega$ given by
\[
\omega = \frac12 d (r^2 \eta)=\frac12 dd^cr^2 \,,
\]
where $d^c=\frac{i}{2}(\bar{\partial}-\partial)$, both $J$ and $\omega$ being
parallel with respect to the Levi-Civita connection $\bar{\nabla}$ of $\bar{g}$.
Moreover, $M$ has odd dimension $2n+1$, where $n+1$ is the complex
dimension of the K\"{a}hler cone. Conversely, given any algebraic K\"{a}hler
orbifold, we can naturally associate to it a quasi-regular Sasakian manifold
\cite{B-G-1999}.

An \emph{Einstein manifold} is a Riemannian manifold
$(M,g)$ satisfying
the Einstein condition
\begin{equation}\label{Einstein}
Ric_{g} = \lambda g \,,
\end{equation}
for some real constant $\lambda$, where $Ric_{g}$ denotes the Ricci
tensor of $g$. Einstein manifolds with $\lambda=0$ are called
\emph{Ricci-flat manifolds}. A \emph{Sasaki-Einstein manifold} is a
Riemannian manifold $(M,g)$ that is both Sasaki and Einstein. We note that
in the case of Sasaki-Einstein manifolds one always has \eqref{Einstein}
with the Einstein constant $\lambda=2n$. We also remark
that Gauss equation relating the curvature of submanifolds
to the second fundamental form shows that a Sasaki manifold $M$ is
Einstein if and only if the metric cone $C(M)$ is  K\"{a}hler Ricci-flat. In particular the
K\"{a}hler cone of an Sasaki-Einstein  manifold has trivial canonical bundle \cite{B-G-2010,Sp}.
We note that one of the most familiar example of homogeneous Sasaki-Einstein
five-manifold is the space $T^{1,1}=S^2\times S^3$ endowed with the following metric
\cite{M-S,C-O}
\[
\begin{split}
ds^2(T^{1,1}) = & \frac16 (d \theta^2_1 + \sin^2 \theta_1 d \phi^2_1 +
d \theta^2_2 + \sin^2 \theta_2 d \phi^2_2) +\\
& \frac19 (d \psi + \cos \theta_1 d \phi_1 + \cos \theta_2 d \phi_2)^2
\,.
\end{split}
\]

A \emph{toric Sasaki manifold} $M$ is a Sasaki manifold
whose K\"{a}hler cone $C(M)$ is a toric K\"{a}hler manifold \cite{Gu}.
In particular, a five-dimensional toric Sasaki-Einstein manifold is a
Sasaki-Einstein manifold with three $U(1)$ isometries: $G=\mathbb{T}^3$. In this
case one can construct canonical coordinates based on the symplectic
geometry of the cone $C(M)$ and specify the Sasaki-Einstein structure
in terms of toric data together with a single function $G$, a symplectic
potential, on the three-dimensional image of the momentum map \cite{O-Y}.
It is known that one of the simplest example of a toric non-orbifold singularity
is the conifold $C(T^{1,1})$, i.e. the Calabi-Yau cone over $T^{1,1}$.

On the other hand, according to Semmelman \cite{Semm}, there is a correspondence
between special Killing forms defined on the Sasaki-Einstein manifold
$M$ and the parallel forms defined on the metric cone $C(M)$.  More exactly, a
$p-$dimensional differential form $\Psi $ is a special Killing form on $M$ if
and only if the corresponding
form
\begin{equation}
\Psi _{cone}:=r^pdr\wedge \Psi +\frac{r^{p+1}}{p+1}d\Psi \,,
\label{eq Semmelmann}
\end{equation}
is parallel on $C(M)$.

In particular, on a five-dimensional Sasaki manifold $M$ with the Reeb
vector field $B$ and $1-$form $\eta := B^\ast$, there are the following two
special Killing forms:
\begin{equation}\label{sKPsi}
\Psi_1 = \eta \wedge d\eta,\quad  \Psi_2=\eta \wedge (d\eta)^2 \,.
\end{equation}
Besides these Killing forms, there are two closed conformal Killing
forms, also called $\ast$-Killing forms, given by
\begin{equation}\label{sKPhi}
\Phi_1 = d\eta,\quad \Phi_2=(d\eta)^2 \,.
\end{equation}
Moreover, in the case of the Calabi-Yau cone $C(M)$ it follows that we have
two additional Killing forms on $M$ connected with the additional parallel
forms of the cone given by the holomorphic complex
volume form $\Omega$ of $C(M)$ and its conjugate \cite{Semm}.

\subsection{Symplectic and complex coordinates on toric manifolds}

Let us consider a toric Sasaki-Einstein manifold. In the spirit of \cite{Abr}
we use in our further considerations the symplectic (action-angle) coordinates
$(y ^i,\Phi ^i)$; here the angular coordinates $\Phi ^i$ will
generate the toric action. The $y^i$ coordinates are obtained using the
momentum map $\mu =\frac 12r^2\eta $, with the correspondence
\begin{equation}
y^i=\mu (\partial /\partial \Phi ^i)\,. \label{xi}
\end{equation}
The K\"ahler form $\omega $ can be written in the simple manner
\cite{Abr, M-S-Y}
\[
\omega =dy^i\wedge d\Phi ^i\,.
\]

In turn, the corresponding K\"ahler metric on the cone $C(M)$ is constructed
using the \emph{symplectic potential} $G$, which is a strictly convex
function $G=G(y)$ of homogeneous degree $-1$ in $y$ \cite{M-S-Y, Abr}. We get
\[
ds^2=G_{ij}dy^idy^j+G^{ij}d\Phi ^id\Phi ^j\,,
\]
where the metric coefficients are computed
\[
G_{ij}=\frac{\partial ^2G}{\partial y^i\partial y^j}\,,
\]
with $\left( G^{ij}\right) =\left( G_{ij}\right) ^{-1}$.

The complex structure $J$ can be described using the above symplectic
coordinates metric coefficients, namely
\[
J=\left(
\begin{array}{cc}
0 & -G^{ij} \\
G_{ij} & 0
\end{array}
\right) \,.
\]

Now we return to the construction of the symplectic potential $G$.

A classical result of Delzant associates to any Delzant polytope $P\in \mathbb{R}^n$
a close connected symplectic manifold $M$, together with a Hamiltonian $\mathbb{T}^n$
action on the manifold, showing that the polytope turns out to be the image
of the momentum map, $P=\mu (M)$. We remaind that a Delzant polytope is a
convex polytope such that there are $n$ edges meeting at each vertex, each
edge meeting at the vertex is of form $1+tu_i$, where $u_i\in \mathbb{Z}^n$, and
$\{u_i\}$ can be chosen to form a basis in $\mathbb{Z}^n$.

A Delzant polytope can be described by the inequalities
\[
l_A(y):=\left\langle y,v_A\right\rangle \ge 0\text{, for }1\le A\le d\,,
\]
where $\{v_A\}$ are inward pointing normal vectors to the facets of the
polytope, $d$ is the number of facets \cite{Abr, Gu}.

\begin{remark}
In the case of the Calabi-Yau cone we take $C(M)$ to be {\it Gorenstein}
which is a necessary condition to admit a Ricci-flat K\"{a}hler metric and $M$
to admit a Sasaki-Einstein metric.
For affine toric varieties it is well-known that $C(M)$ being Gorenstein is equivalent
to the existence of a basis for the torus $\mathbb{T}^n$ for which
\begin{equation}\label{Gorenstein}
v_i=(+1,w_i)\,,
\end{equation}
for each $a,\cdots ,d$ and $w_a \in\mathbb{Z}^{n-1}$ \cite{M-S,M-S-Y}.
\end{remark}

If $B$ is the Reeb vector field, let us point out the following relation
which links this geometric object to the metric coefficients
\[
B_i=2G_{ij}y^j.
\]
Now let us also define the affine function $l_B:=\left\langle B,\cdot
\right\rangle $, and $l_\infty :=\left\langle \sum_Av_A,\cdot \right\rangle $.
Then, the symplectic potential $G$ can be written in terms of the toric
data \cite{Abr,Gu}
\begin{equation}
G=G^{can}+G^B+h,  \label{G}
\end{equation}
where
\begin{eqnarray}
G^{can} &=&\frac 12\sum_Al_A(y )\log l_A(y),  \label{G_can_G_B} \\
G^B &=&\frac 12l_B(y)\log l_B(y)-\frac 12l_\infty (y)\log l_\infty (y)\,,
\nonumber
\end{eqnarray}
and $h$ is a homogeneous function of degree $1$ in variables $y$.
In the general case, as $G$ needs to satisfy the
Monge-Amp\`ere equation, the function $h$ is added.

For a complete determination of the symplectic potential $G$ it is necessary
to compute the Reeb vector $B$ and the function $h$. There are two known
different algebraic procedures to extract the components of the Reeb vector
from the toric data. According to the AdS/CFT correspondence the volume of
the Sasaki-Einstein space corresponds to the central charge of the dual
conformal field theory. The first procedure is based on \emph{the maximization of
the central charge} ($a$-maximization) \cite{Int-Wec} used in connection with
the computation of the Weyl anomaly in 4-dimensional field theory. The
second one is known as \emph{volume minimization} (or $Z$-minimization) \cite{M-S-Y}.

The symplectic potential allow us to pass to the coordinate patch $(x^i,\Phi^i)$
obtained from complex coordinates $z^i:=x^i+\mathrm{i}\Phi ^i$, with
$\mathrm{i}:=\sqrt{-1}$; this is possible via the Legendre transform which
relates the symplectic potential $G$ and the K\"ahler potential $F$
\[
F(x)=\left( y^i\frac{\partial G}{\partial y^i}-G\right) :
\left(y=\partial F/\partial x\right) \,.
\]

Consequently $F$ and $G$ are Legendre dual to each other
\[
F(x)+G(y)=\sum_j\frac{\partial F}{\partial x^j}\frac{\partial G}{\partial y^i}\,\,,
\]
and
\[
x^i=\frac{\partial G}{\partial y^i},\mbox{\,\,}y^i=\frac{\partial F}{\partial x^i}\,.
\]

The metric structure is now written in the following manner
\[
ds^2=F_{ij}dx^idx^j+F_{ij}d\Phi ^id\Phi ^j\,,
\]
where the metric coefficients are again obtained using the Hessian of the
K\"ahler potential $F$, i.e.
\[
F_{ij}=\frac{\partial ^2F}{\partial x^i\partial x^j}\,.
\]
Note also that $\left( F_{ij}\right) =\left( G^{ij}\right) $ \cite{Abr}.

With respect to the coordinates $(x^i,\Phi ^i)$, the K\"ahler form is
\[
\omega =\left(
\begin{array}{cc}
0 & F_{ij} \\
-F_{ij} & 0
\end{array}
\right) \,.
\]

We use in the following the fact that the Calabi-Yau metric cone is Ricci
flat. From the classical formula
\[
\rho =-\mathrm{i}\partial \bar \partial \log \det (F_{ij})\,,
\]
we get
\begin{equation}
\det (G_{ij})=\exp{\left(2\gamma _i\frac{\partial G}{\partial y^i}-c\right)}\,,
\label{det_G}
\end{equation}
with constants $\gamma _i$, and $c$. Using \eqref{G}-\eqref{G_can_G_B}, we are
able to express the coordinates $x^i$ and the metric coefficients $G_{ij}$
\[
\begin{split}
x^i &=\frac{\partial G}{\partial y^i}=\frac 12\sum_Av_A^i\log l_A(y)
+\frac 12B^i(1+\log l_B(y)) \\
&-\frac 12\sum_Av_A^i\log l_\infty (y)+\lambda _i   \,,\\
G_{ij} &=\frac 12\sum_A\frac{v_A^iv_A^j}{l_A(y)}
+\frac 12\frac{B_iB_j}{l_B(y)}
-\frac 12\frac{\sum_Av_A^i\sum_Av_A^j}{l_\infty (y)}\,.
\end{split}
\]

Now, as the metric has to be smooth, from (\ref{det_G}) it turns out that \cite{O-Y}
\[
\gamma =(-1,0,..,0) \,,
\]
and
\[
\det (F_{ij})=\exp{\left(2x^1+c\right)}.
\]

If $\mathrm{Vol}$ is the volume form on the metric cone, then the holomorphic volume
form $\Omega$ satisfies
\[
\mathrm{Vol}=\frac{i^{n+1}}{2^{n+1}} (-1)^{n(n+1)/2}\Omega \wedge \bar{\Omega}
=\frac{1}{(n+1)!} \omega ^{n+1}\,.
\]

Then, eventually ignoring the multiplicative constant, in complex coordinates
$\Omega $ can be written as \cite{M-S-Y}
\begin{eqnarray*}
\Omega  &=&\exp({i\alpha })\det (F_{ij})^{1/2}dz^1\wedge ..\wedge dz^n \\
&=&\exp({x^1+i\alpha})dz^1\wedge ..\wedge dz^n\,.
\end{eqnarray*}
As $\Omega $ is parallel, it is also closed. Then we
can fix the phase $\alpha $ to be $\Phi ^1$, and  we obtain the following
simple formula for the holomorphic volume form, which motivates the interest
for complex (and consequently, symplectic) coordinates

\begin{equation}
\Omega =\exp (z^1)dz^1\wedge ..\wedge dz^n.  \label{Omega}
\end{equation}
Employing the above relation, in the next sections we show that it is possible
to extract the special Killing forms on manifolds of Sasaki-Einstein type.

\section{Symplectic and complex coordinates on conifold $C(T^{1,1})$}

Throughout this section we introduce complex coordinates on $C(T^{1,1})$ using the
classical procedure exposed above.

We start out by considering the global defined contact 1-form
\begin{equation}
\eta =\frac 13(d\psi +\cos \theta _1d\phi _1+\cos \theta _2d\phi _2)\,.
\label{eta}
\end{equation}

This form allows us to
construct on $C(T^{1,1})$ the symplectic form (see e.g. \cite{M-S})
\[
\begin{split}
\omega & =-\frac{r^2}6(\sin \theta _1d\theta _1\wedge d\phi_1
+\sin \theta_2d\theta _2\wedge d\phi _2) \\
& +\frac 13rdr\wedge (d\psi +\cos \theta _1d\phi _1+\cos \theta _2d\phi_2)\,.
\end{split}
\]

Furthermore, if we employ the basis \cite{M-S} for an effectively acting
$\mathbb{T}^3$ action
\[
\begin{split}
e_1& =\frac \partial {\partial \phi _1}+\frac 12\frac \partial {\partial \nu}\,, \\
e_2& =\frac \partial {\partial \phi _2}+\frac 12\frac \partial {\partial \nu}\,, \\
e_3& =\frac \partial {\partial \nu }\,,
\end{split}
\]
where $2\nu =\psi $, then, considering action coordinates associated with this basis,
we get the momentum map using \eqref{xi} (see also \cite{M-S})
\[
\mu =\left( \frac 16r^2(\cos \theta _1+1),\frac 16r^2(\cos \theta_2+1),\frac 13r^2\right)
\,.  \label{mom}
\]

The Reeb vector field $B$ has the form
\begin{equation}\label{Reeb}
B = 3 \frac{\partial}{\partial \psi} = \frac{3}{2} \frac{\partial}{\partial \nu}\,,
\end{equation}
and is easy to see that $\eta(B)=1$.

Now let us consider the ``inward pointing'' primitive normal vectors to the cone
\[
v_1^{\prime}=[-1,0,1]\,,\,v_2^{\prime}=[0,-1,1]\,,\,v_3^{\prime}=[1,0,0]\,
\,,\,v_4^{\prime}=[0,1,0]\,.  \label{v'}
\]

We apply a $SL(3;\mathbb{Z})$ transformation $T$ (see also \cite{M-S})
\[
T=\left(
\begin{array}{ccc}
1 & 1 & 2 \\
0 & 1 & 1 \\
0 & 0 & 1
\end{array}
\right) \,,
\]
to bring the vectors $v_i=Tv_i^{\prime }$  in the form \eqref{Gorenstein} \cite{O-Y}
\begin{equation}
v_1=[1,1,1]\,,\,v_2=[1,0,1]\,,\,v_3=[1,0,0]\,,\,v_4=[1,1,0]\,. \label{v}
\end{equation}

According to the above transformation we obtain the new basis
\begin{equation}\label{eprime}
\begin{split}
e_1^{\prime }& =\frac \partial {\partial \phi _1}+\frac 12\frac \partial{\partial \nu }\,, \\
e_2^{\prime }& =\frac \partial {\partial \phi _2}-\frac \partial {\partial\phi _1}\,, \\
e_3^{\prime }& =-\frac \partial {\partial \phi _1}-\frac \partial {\partial\phi _2}\,,
\end{split}
\end{equation}
where $e_i^{\prime }:=e_iT^{-1}$.

We consider the new angle coordinates
\[
\begin{split}
\Phi ^1& :=2\nu =\psi \,, \\
\Phi ^2& :=-\frac 12\phi _1+\frac 12\phi _2+\nu =-\frac 12\phi _1+\frac12\phi _2
+\frac 12\psi,\\
\Phi ^3& :=-\frac 12\phi _1-\frac 12\phi _2+\nu =-\frac 12\phi _1-\frac12\phi _2
+\frac 12\psi\,.
\end{split}
\]
and it is easy to check that
\[
e_i^{\prime }=\frac \partial {\partial \Phi ^i}\,,
\]

In this new basis, applying \eqref{xi}, the momentum map becomes:
\begin{equation}
\mu ^{\prime }= y =\left( \frac 16r^2(\cos \theta _1+1),
\frac16r^2(\cos \theta _2-\cos \theta _1),
-\frac 16r^2(\cos \theta _1+\cos \theta_2)\right)\,.
\end{equation}

This way we end up with the symplectic action-angle coordinates $(y^i$, $\Phi ^i)$,
for $1\le i\le 3$.

Now, in order to introduce the complex coordinates on conifold we need the
symplectic potential $G$. In the particular case of the conifold $T^{1,1}$, the sum
$G^{can}+G^B$ is already a solution of the Monge-Amp\`ere equation. Consequently the
function $h$ is not needed anymore, and the equations \eqref{G}-\eqref{G_can_G_B}
simplify as (see e.g. \cite{O-Y})
\begin{equation}\label{G_conifold}
G=G^{can}+G^B\,,
\end{equation}
where
\[
G^{can}=\sum_{A=1}^4\frac 12\left\langle v_A,y \right\rangle \log\left\langle v_A,y
\right\rangle \,,
\]
and
\[
G^B=\frac 12\left\langle B,y \right\rangle \log \left\langle B,y
\right\rangle -\frac 12\left\langle B^{can},y \right\rangle \log \left\langle
B^{can},y \right\rangle \,.
\]

In the above relation the vectors $v_A$ are just \eqref{v} and
\[
B^{can}=\sum_1^4v_A=(4,2,2)\,.
\]

Concerning the Reeb vector field on $T^{1,1}$ \eqref{Reeb} written in the
new basis $\{e_i^{\prime}\}$ \eqref{eprime}, it has the components
\[
B=(3,3/2,3/2)\,,
\]
consistent with the determination from toric data using $Z$-minimization  \cite{Int-Wec}
or $a$-maximization \cite{M-S-Y}.

We construct complex coordinates using \eqref{G_conifold} and the Legendre transform
\[
\begin{split}
x^i=\frac{\partial G}{\partial y^i}=& \frac 12\sum_1^4 v_{A}^{i}\log
<v_A,y >+\frac 12B^i\log <B, >+\frac 12B^i \\
& -\frac 12(B^{can})^i\log <B^{can},y >\,.
\end{split}
\]

But it is easy to see that
\begin{equation}
\begin{split}
\sum_1^4 v_{A}^{1}\log <v_A,y >=& 8\log r+2\log \sin \theta _1+2\log \sin
\theta _2 \\
& -4\log 2-4\log 3, \\
\sum_1^4 v_{A}^{2}\log <v_A,y >=& 4\log r+\log (1-\cos \theta _1)+\log
(1+\cos \theta _2) \\
& -2\log 2-2\log 3, \\
\sum_1^4 v_{A}^{3}\log <v_A,y >=& 4\log r+\log (1-\cos \theta _1)+\log
(1-\cos \theta _2) \\
& -2\log 2-2\log 3, \\
\log <B,y >=& 2\log r-\log 2, \\
\log <B^{can},y >=& 2\log r+\log 2-\log 3\,.
\end{split}
\label{g1}
\end{equation}

Using now (\ref{g1}) and basic trigonometric formulas, we derive
\[
\begin{split}
x^1& =3\log r+\log \sin \theta _1+\log \sin \theta _2+\frac 32-\frac{11}
2\log 2\,, \\
x^2& =\frac 32\log r+\log \sin \frac{\theta _1}2+\log\cos\frac{\theta _2}2
+\frac 34-\frac{11}4\log 2\,, \\
x^3& =\frac 32\log r + \log \sin \frac{\theta _1}2+\log\sin\frac{\theta_2}2
\frac 34-\frac{11}4\log 2\,.
\end{split}
\]

In the sequel, for the sake of simplicity we will ignore the additive constants.
Therefore, in accordance with \cite{M-S-Y} we can introduce on conifold $C(T^{1,1})$
the following patch of complex coordinates
\begin{equation}
\label{Z}
\begin{split}
z^1 &=3\log r+\log \sin \theta _1+\log \sin \theta _2+\mathrm{i}\psi , \\
z^2 &=\frac 32\log r+\log \sin \frac{\theta _1}2+\log \cos \frac{\theta _2}2\,   \\
&+\frac{\mathrm{i}}2(\psi +\phi _1+\phi _2),   \\
z^3 &=\frac 32\log r+\log \sin \frac{\theta _1}2+\log \sin \frac{\theta _2}2\,   \\
&+\frac{\mathrm{i}}2(\psi -\phi _1-\phi _2)\,.
\end{split}
\end{equation}

Now, regarding \eqref{Omega}, we see that the above coordinates are precisely the necessary ingredient in order to
extract the special Killing forms on our homogeneous Sasaki-Einstein manifold.

\section{Special Killing forms on $T^{1,1}$}

Applying the above general results, in this section we obtain the complete set of
special Killing forms on the manifold $T^{1,1}$.

First of all, we have to calculate the holomorphic volume form $\Omega $.
Starting out with \eqref{Z}, we obtain
\[
\begin{split}
& \exp( {z^1})=r^3\sin \theta _1\sin \theta _2\exp \mathrm{i}{\psi }\,,
\\
& dz^1=\frac 3rdr+T_1, \\
& dz^2=\frac 3{2r}dr+T_2\,, \\
& dz^3=\frac 3{2r}dr+T_3\,.
\end{split}
\]
where
\begin{equation}\label{T}
\begin{split}
T_1&:=\cot \theta _1d\theta _1+\cot \theta _2d\theta _2+\mathrm{i}d\psi ,\\
T_2 &:=\frac 12\cot \frac{\theta_1}{2} d\theta _1-\frac 12\tan \frac{\theta _2}2d\theta_2
+\frac{\mathrm{i}}2(d\psi -d\phi _1+d\phi _2), \\ 
T_3 &:=\frac 12\cot \frac{\theta_1}{2} d\theta _1+\frac 12\cot \frac{\theta _2}2d\theta_2
+\frac{\mathrm{i}}2(d\psi -d\phi _1-d\phi _2)\,. 
\end{split}
\end{equation}

Now we calculate the holomorphic volume form (see e.g. \cite{M-S-Y})
\begin{equation}
\begin{split}
\Omega &=\exp (z^1)dz^1\wedge dz^2\wedge dz^3 \\
\ &=\exp (z^1)(\frac 3rdr+T_1)\wedge (\frac 3{2r}dr+T_2)\wedge (\frac
3{2r}dr+T_3) \,. \nonumber
\end{split}
\end{equation}

In our particular framework the equation \eqref{eq Semmelmann} becomes
\[
\Omega =r^2dr\wedge \Psi +\frac{r^3}3d\Psi \,.
\]

In order to extract $\Psi$ we have to keep the trace of the differential
form $dr$ in the above equation. We clearly get

\begin{equation}
\Psi =3\sin \theta _1\sin \theta _2 \exp({\mathrm{i}\psi })(\frac 14T_2\wedge
T_3-\frac 12T_1\wedge T_3+\frac 12T_1\wedge T_2)\,.  \label{TT}
\end{equation}
We calculate the above wedge products using \eqref{T}. For the first wedge product in
\eqref{TT}, after calculations we get

\begin{equation}
\begin{split}
T_2\wedge T_3 &=\frac 12\cot \frac{\theta _1}2\frac 1{\sin \theta_2}d\theta _1\wedge d\theta _2-
\frac{\mathrm{i}}2\cot \frac{\theta _1}2d\theta _1\wedge d\phi _2  \label{T2T3} \\
&\ -\frac{\mathrm{i}}2\frac 1{\sin \theta _2}d\theta _2\wedge d\psi +\frac
{\mathrm{i}}{2}\frac 1{\sin \theta _2}d\theta _2\wedge d\phi _1-\frac{\mathrm{i}}2\cot
\theta _2d\theta _2\wedge d\phi _2   \\
&\ -\frac 12d\phi _1\wedge d\phi _2+\frac 12d\psi \wedge d\phi _2\,.
\end{split}
\end{equation}
For the second product we obtain

\begin{equation}
\begin{split}
T_1\wedge T_3 &=\frac 12(\cot \theta _1\cot \frac{\theta _2}2
-\cot \theta_2\cot \frac{\theta _1}2)d\theta _1\wedge d\theta_2
-\frac{\mathrm{i}}2\frac1{\sin \theta _1}d\theta _1\wedge d\psi  \label{T1T3} \\
&-\frac{\mathrm{i}}2\cot \theta _1d\theta _1\wedge d\phi_1
-\frac{\mathrm{i}}2\cot \theta _1d\theta _1\wedge d\phi _2
-\frac{\mathrm{i}}2\frac 1{\sin\theta _2}d\theta _2\wedge d\psi   \\
&-\frac{\mathrm{i}}2\cot \theta _2d\theta _2\wedge d\phi _1
-\frac{\mathrm{i}}2\cot \theta _2d\theta _2\wedge d\phi _2+\frac 12d\psi \wedge d\phi_1
+\frac 12d\psi \wedge d\phi _2\,.
\end{split}
\end{equation}
Finally, for the last product we get

\begin{equation}
\begin{split}
\label{T1T2}
T_1\wedge T_2 &=-\frac 12(\cot \theta _1\tan \frac{\theta _2}2
-\cot \theta_2\cot \frac{\theta _1}2)d\theta _1\wedge d\theta _2
-\frac{\mathrm{i}}2\frac1{\sin \theta _1}d\theta _1\wedge d\psi   \\
&\ \ -\frac{\mathrm{i}}2\cot \theta _1d\theta _1\wedge d\phi_1
+\frac{\mathrm{i}}2\cot \theta _1d\theta _1\wedge d\phi_2
+\frac{\mathrm{i}}2\frac1{\sin \theta _2}d\theta _2\wedge d\psi   \\
&\ -\frac{\mathrm{i}}2\cot \theta _2d\theta _2\wedge d\phi_1
+\frac{\mathrm{i}}2\cot \theta _2d\theta _2\wedge d\phi _2+\frac 12d\psi \wedge d\phi_1
-\frac 12d\psi \wedge d\phi _2\,.
\end{split}
\end{equation}

Now we plug \eqref{T2T3}-\eqref{T1T2} in \eqref{TT} and we end up with
the simple formula bellow for the special complex Killing form
(in what follows we ignore the multiplicative constants)
\[
\begin{split}
\Psi & =\exp({\mathrm{i}\psi})  \left[ 2 d\theta_1\wedge d\theta_2
-2\mathrm{i}\sin \theta_2 d\theta_1\wedge d\theta_2 \right. \\
& \left. ~~~~+2\mathrm{i}\sin \theta_1 d\theta_2\wedge d\phi_1
-2\sin \theta_1 \sin \theta_2 d\phi _1\wedge d\phi_2 \right] \,.
\end{split}
\]

From here, we can easily derive the real special Killing forms computing the real and
imaginary part of $\Psi $:
\[
\begin{split}
\Re \Psi & =\cos \psi \,d\theta _1\wedge d\theta _2+\sin \theta _2\sin \psi
\,d\theta _1\wedge d\phi _2 \\
& ~~-\sin \theta _1\sin \psi \,d\theta _2\wedge d\phi _1-\sin \theta _1\sin
\theta _2\cos \psi \,d\phi _1\wedge d\phi _2\,,
\end{split}
\]
\[
\begin{split}
\Im \Psi & =\sin \psi \,d\theta _1\wedge d\theta _2-\sin \theta _2\cos \psi
\,d\theta _1\wedge d\phi _2 \\
& ~~+\sin \theta _1\cos \psi \,d\theta _2\wedge d\phi _1-\sin \theta _1\sin
\theta _2\sin \psi \,d\phi _1\wedge d\phi _2\,.
\end{split}
\]

Finally, we calculate the Killing forms $\Phi _1$ and $\Phi _2$ \eqref{sKPhi},
$\Psi _1$ and $\Psi _2$ \eqref{sKPsi} using the contact 1-form $\eta $ \eqref{eta}.
We obtain
\[
\Phi _1=d\eta =-\frac 13(\sin \theta _1d\theta _1\wedge d\phi _1+
\sin \theta_2d\theta _2\wedge d\phi _2)\,,  \label{1}
\]
\[
\Phi _2=(d\eta )^2=-\frac 29\sin \theta _1\sin \theta _2d\theta _1\wedge
d\theta _2\wedge d\phi _1\wedge d\phi _2 \,, \label{2}
\]
and respectively
\[
\begin{split}
\Psi _1 =&\eta \wedge d\eta =\frac 19(\sin \theta _1d\psi \wedge d\theta
_1\wedge d\phi _1+\sin \theta _2d\psi \wedge d\theta _2\wedge d\phi _2  \\
& -\cos \theta _1\sin \theta _2d\theta _2\wedge d\phi _1\wedge d\phi
_2+\cos \theta _2\sin \theta _1d\theta _1\wedge d\phi _1\wedge d\phi _2)\,,\\
\Psi _2=&\eta \wedge (d\eta )^2=-\frac 2{27}\sin \theta _1\sin \theta _2d\psi
\wedge d\theta _1\wedge d\theta _2\wedge d\phi _1\wedge d\phi _2 \,.
\end{split}
\]

\section{Conclusions}

The significance of the present work relies on unquestionable relevance of
conformal Killing-Yano tensors in both mathematics and physics. It is known
that it is a very difficult problem to find solutions of the conformal Killing-Yano
equations on arbitrary Riemannian manifolds, but fortunately, in the case of
spaces endowed with remarkable geometrical structures the explicit construction
of the Killing forms is permitted.

In this paper we have obtained the complete set of special Killing forms on the
five-dimensional Sasaki-Einstein space $T^{1,1}$ with an approach based on the
interplay between symplectic and complex coordinates on K\"{a}hler toric manifolds.
On the other hand, concerning the potential of the present paper, as a lot of
non-trivial examples of toric Sasaki-Einstein manifolds occurs in the recent
literature (see, e.g., \cite{CFO,MS,VC}), it is both natural and useful to extend
the present work to other spaces of interest. In fact, using toric geometry many
examples of Sasaki-Einstein manifolds can be constructed, and these spaces are a
good testing ground for the predictions of the AdS/CFT correspondence \cite{MSY}.

\section*{Acknowledgments}
MV was supported by CNCS-UEFISCDI, project number PN-II-ID-PCE-2011-3-0137.
The work of GEV  was supported by CNCS-UEFISCDI, project number PN-II-ID-PCE-2011-3-0118.

\end{document}